\documentclass[prl,showpacs,preprintnumbers,twocolumn]{revtex4}
\usepackage{graphicx}
\begin{document}

\title{Closed-circuit domain quadruplets in BaTiO$_3$ nanorods embedded in SrTiO$_3$ film}

\author{ V. Stepkova$^{\rm a}$}
\author{ P. Marton$^{\rm a}$}
\author{ N. Setter$^{\rm b}$}
\author{ J. Hlinka$^{\rm a}$}
\affiliation{$^{\rm a}${\em{Institute of Physics, Academy of Sciences of the Czech Republic, Na Slovance 2, 18221 Praha 8, Czech Republic}}\\
$^{\rm b}${\em{Ceramics Laboratory, Swiss Federal Institute of Technology, Lausanne CH-1015, Switzerland}}\\
}
\date{\today}

\begin{abstract}
Cylindrical BaTiO$_3$ nanorods embedded in $\langle 100 \rangle $-oriented SrTiO$_3$ epitaxial film in a brush-like configuration are investigated  in the framework of the
Ginzburg-Landau-Devonshire model. It is shown that strain compatibility at BaTiO$_3$/SrTiO$_3$ interfaces keeps BaTiO$_3$ nanorods in the rhombohedral phase even at room temperature.  Depolarization field at the BaTiO$_3$/SrTiO$_3$ interfaces  is  reduced by an emission of the 109-degree or 71-degree domain boundaries.
In case of 10-80\,nm diameter nanorods,  the ferroelectric domains are found to form a quadruplet with a robust flux-closure arrangement of the in-plane components of the spontaneous polarization.
 The out-of-plane components of the polarization are either balanced or oriented up or down along the nanorod axis.
  Switching of the out-of-plane polarization with coercive field of about $5 \times 10^6$\,V/m occurs as a collapse of a 71-degree cylindrical domain boundary formed at the curved circumference surface of the nanorod. The remnant domain quadruplet configuration is chiral, with the $C_4$ macroscopic symmetry. More complex stable domain configurations with coexisting clockwise and anticlockwise quadruplets contain interesting arrangement of strongly curved 71-degree boundaries.
\end{abstract}

%\begin{keywords}
%\end{keywords}

\pacs{77.80.-e,77.80.Dj,77.84.-s}

% 77.80.-e Ferroelectricity and antiferroelectricity
% 77.22.Ej  Polarization and depolarization
% 77.80.Dj  Domain structure; hysteresis (for domain structure and hysteresis in ferromagnetic materials, see 75.60.-d)
% 77.80.Fm  Switching phenomena (for ultrafast magnetization dynamics and switching, see 75.78.Jp; for spintronics, see 85.75.-d)
% 77.65.-j  Piezoelectricity and electromechanical effects
% 68.37.-d  Microscopy of surfaces, interfaces, and thin films
% 68.37.Ps  Atomic force microscopy (AFM)
% 68.55.-a  Thin film structure and morphology (for methods of thin film deposition, film growth and epitaxy, see 81.15.-z)
% 77.84.Bw  Elements, oxides, nitrides, borides, carbides, chalcogenides, etc.
% 77.84.-s  Dielectric, piezoelectric, ferroelectric, and antiferroelectric materials (for nonlinear optical materials, see 42.70.Mp; for dielectric materials in electrochemistry, see 82.45.Un)

\maketitle
%----------------------------------------------------------------------------------

Ferroelectric objects with nanoscale dimensions, such as  nanodots, nanorods, nanotubes and nanolayers, both free-standing and embedded in thin films and bulk composites, are generally believed to provide a range of very attractive physical properties.\cite{Scott06Rev,Lee,Cata13,Slutsker2008} Recent progress in this field is remarkable -- free-standing  single crystal ferroelectric nanorods  have been  prepared from a number of classical ferroelectric materials\cite{Mao,Urban,Yun2002,Saeterli2010,Vasc05A, Luo2003, Wang12B,Spanier2006} and their properties measured,\cite{Suya04D,Wang08F,Yama10H,Spanier2006,ZWang} growing of planar ferroelectric superlattices has become a well established approach for material property design through the epitaxial strain,\cite{Schlom,Waru,Lee05, Slutsker2006} self-assembled 1-3 ferroelectric-paralectric nanocomposites were used to tailor the dielectric tunability\cite{Yamada2009}  and the ultrafast switching demonstrated for nanoscale  capacitors already opens new perspectives for practical devices.\cite{Scott06Rev,Li,Gruverman2008A,Gruverman2008,Gruverman2009,Jung, Fridkin,Kim}  Properties of all these ferroelectric nanoelements are strongly sensitive to the conditions of their surface.\cite{Slutsker2008,Wang2011} Uncompensated polarization charges at the surface perpendicular to the  spontaneous polarization can lead to a strong modification of the ferroelectric state or even completely suppress the ferroelectricity,\cite{Ondr13} and even more complex behavior can be expected in composite nanostructures due to the epitaxial strains at the material interfaces.

Moreover, the established scaling laws for domain sizes suggest that ferroelectric nanoelements may host fairly small ferroelectric domains, that in turn can play a decisive role in determining their properties.\cite{Schilling2006,Schilling2009} In particular, numerous previous theoretical simulations have reported various closed-circuit domain configurations and vortices in electrically isolated ferroelectric nanodots and nanorods.\cite{Fu, Naumov, Stachiotti, Baudry, Baudry2012, Chen2012, Laho08} Interestingly,  the straightforward experimental evidence for these very interesting natural topological defects is so far very limited and quite often, an alternative star-like "quadrant-quadrupole" arrangement\cite{Schilling,Gregg} was observed in tetragonal ferroelectrics instead of the "flux-closure" curling polarization state.\cite{Schilling,Boro13,Ahlu13,McQuaid,Gregg,Chang,Ivry,McGilly2010} Considerations about possible ways to realize such curling polarization states have led us to theoretical investigations of the behavior of a model system consisting of ferroelectric BaTiO$_3$ nanorods embedded in a matrix of a thin dielectric film of insulating SrTiO$_3$. The aim of this paper is to report the condition of formation of these  peculiar closed-circuit domain configurations in this system.

% here  Figure 1
\begin{center}
\begin{figure}
\includegraphics[width=6.5cm]{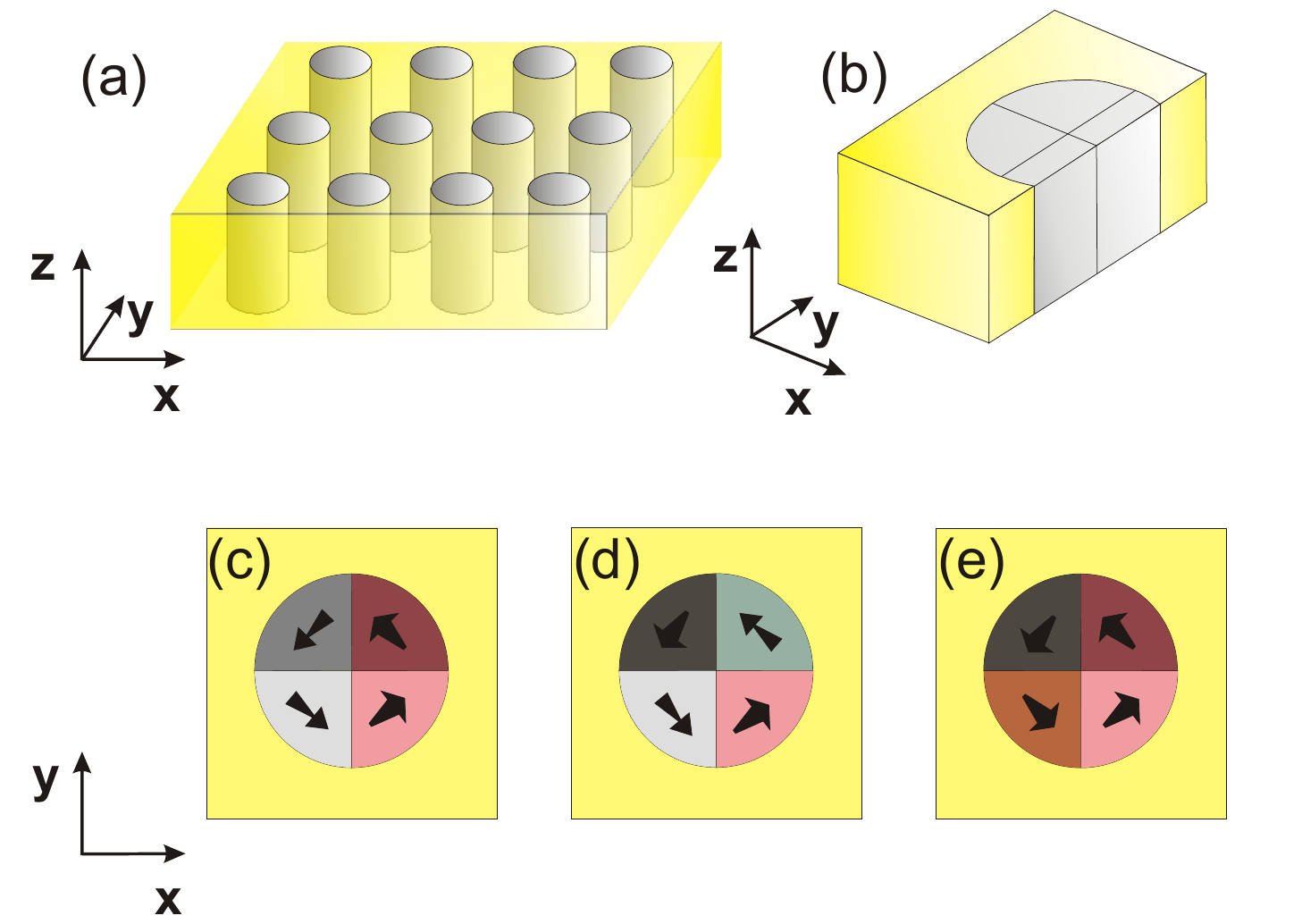}
\caption{(Color online) Scheme of the investigated heterostructure and its nanodomain states: (a) brush-like arrangement of the BaTiO$_3$ nanorods embedded epitaxially in the $[001]$ oriented SrTiO$_3$ epitaxial film;
(b) domain boundaries within the 40\,nm nanorod, (c) stable domain configuration in the 40\,nm nanorod, $z$-components of polarization arranged in a up-up-down-down manner;
(d)  another stable domain structure, with
$z$-components of polarization arranged in a up-down-up-down manner, (e) still another stable arrangement, with
$z$-components of polarization arranged in a up-up-up-up manner. All calculations assumed 3D periodic boundary conditions.
} \label{structure}
\end{figure}
\end{center}

Phase-field simulations presented in this article are based on the
phenomenological Ginzburg-Landau-Devonshire (GLD) model \cite{my,Hlin09,Mart10} in which the excess
Gibbs free-energy functional $F$ is expressed in terms  of ferroelectric polarization $P_i$, its
spatial derivatives $P_{i,j}=\partial P_{i}/\partial x_{j} $ and
strain components $e_{ij}=(\partial u_{i}/
\partial x_{j} + \partial u_{j}/ \partial x_{i})/2$,  $i,j=1-3$ as
\begin{eqnarray}
F= \int d{\rm {\bf r}} f_{\rm GLD} [\{P_i,P_{i,j},e_{ij} \}] +
F_{\rm dip}[\{P_i\}]~,\label{eqn1}
\end{eqnarray}
where the first term
\begin{eqnarray}\label{eqn_total_potential}
f_{\rm GLD} [\{P_i,P_{i,j},e_{ij} \}]= \nonumber \\
 \alpha_1 \sum_i P_i^2+\alpha_{11}^{\rm (e)}\sum_i P_i^4+
\alpha_{12}^{\rm (e)}\sum_{i>j} P_i^2P_j^2\nonumber \\
+ \alpha_{111}\sum_i P_i^6+\alpha_{112}\sum_{i>j}
(P_i^4P_j^2+P_j^4P_i^2) \nonumber \\
 + \alpha_{123}P_1^2P_2^2P_3^2
 -P_i E_i - \sigma_{ij}  e_{ij}\nonumber \\+ \frac{1}{2}C_{ijkl}e_{ij}e_{kl} -q_{ijkl}e_{ij}P_k P_l +
 \frac{1}{2}G_{ijkl}P_{i,j}P_{k,l}.\label{GLD}
\end{eqnarray}
stands for the room-temperature GLD functional, $C_{ijkl}, q_{ijkl}$ and $G_{ijkl}$ stand for
components of the elastic, electrostriction and gradient tensors, while
$E_i$ and $\sigma_{ij}$ stands for the components of
  external electric field and the applied homogeneous stress tensor, respectively.
The other term in Eq.\,(\ref{eqn1}) describes the electrostatic
energy\cite{HuChen3D,Khacha}
\begin{eqnarray}
F_{\rm dip}[\{P_i\}]=-\frac{1}{2}\int{ d{\rm {\bf r}} [{\bf
E}_{\rm dip} ({\bf r}) \cdot {\bf P}({\bf r})]},
\end{eqnarray}
associated with the  interaction of individual dipoles
with the  field ${\bf E}_{\rm dip}$ created by
the inhomogeneous part of the polarization field
\begin{eqnarray}
  {\bf E}_{\rm dip}({\bf r}) =  \frac{-1}{4 \pi \epsilon_0 \epsilon_{\rm
B} } \int d{\rm {\bf r'}} \left[\frac{{\bf P}({\bf r'})}{|{\bf
R}|^3}
 -\frac{3({\bf P}({\bf r'})\cdot{\bf
R}) ~ {\bf R}}{|{\bf R}|^5}\right] , \label{dipdip}
\end{eqnarray}
where ${\bf R}={\bf r}-{\bf r'}$, $\epsilon_{\rm B}$ is the
relative background permittivity of the medium (without the
primary order parameter contribution) and $\epsilon_{\rm 0}$ is
the permittivity of vacuum.

In order to find the equilibrated domain structure under specified
conditions, we have applied the usual phase-field approach
\cite{Nambu,HuChen3D,Artemev,simul,Slutsker2008} consisting in
simulation of the natural "equilibration" process by numerical
solution of the corresponding time-dependent Ginzburg-Landau
equation for the field of polarization
\begin{eqnarray}
  \frac{\partial P_i}{\partial t} = -  \Gamma \frac{\delta F}{\delta P_i}
\label{GLeq}
\end{eqnarray}
where $\Gamma$ is a kinetic coefficient controlling the energy
dissipation rate of the system. Mechanical equilibrium is assumed
to be achieved at each instant so that the inhomogeneous
strain field can be eliminated from the energy functional of
Eq.\,(\ref{eqn1}) using the corresponding Euler-Lagrange
equations.\cite{Nambu,Khacha,simul,Klot03}
GLD model parameters for the room temperature BaTiO$_3$ were selected same as in Refs.\,\onlinecite{Hlin09, Ondr13}.
For SrTiO$_3$, we have used free-energy coefficients of Ref.\,\onlinecite{Sheng}.
The most essential difference at room temperature is obviously the positive value of the quadratic coefficient $\alpha_{1}$ in SrTiO$_3$. Since SrTiO$_3$ and BaTiO$_3$ are rather similar materials, the parameters defining the gradient, elastic, electrostrictive and dipole-dipole interactions  of SrTiO$_3$ were taken same as for BaTiO$_3$.\cite{Yama72} Having in mind an  epitaxial film clamped to a substrate with an effective lattice constant comparable to cubic BaTiO$_3$, we have imposed the average in-plane strain components and out-of-plane stress components to zero ($\langle e_{11} \rangle = \langle e_{22} \rangle = \langle e_{12} \rangle =0$, $\sigma_{33}=\sigma_{23}=\sigma_{13}=0$). To define a realistic time
scales in the simulated processes, the kinetic coefficient was set to $\Gamma = 4\times\rm10^4\,C^2 J^{-1} m^{-1} s^{-1}$, in agreement with the earlier estimations \cite{timescale}.
The simulations were conducted under periodic boundary conditions within a 2D or 3D rectangular discrete arrays.
The equation (\ref{GLeq}) was resolved
numerically in Fourier space by a second-order semi-implicit
method with spatial steps 0.5\,nm and individual time steps 0.5\,fs.

The investigated nanostructure is sketched in Fig.\,\ref{structure}a. Since the nanorods are fully embedded into the SrTiO$_3$ matrix,  the epitaxial matching prevents the development of  large electrostrictive distortion in the BaTiO$_3$ nanorods, and, consequently, it strongly influences the competition between various ferroelectric states.  Nanorods of 10-80\,nm diameter studied in the present work were in the rhombohedral ferroelectric state, with local polarization close to the $\langle 111 \rangle$ directions. To minimize the depolarization field, the polarization tends to be tangential to the BaTiO$_3$/SrTiO$_3$ interface. Stable configurations obtained for the 10-80\,nm  diameter nanorods typically break up in 4 domains, separated by 2 perpendicular planar domain boundaries intersecting on the axis of the rod (Fig.\,\ref{structure}b). The adjacent domains are arranged in a head-to-tail manner, so that the  in-plane polarization components are forming a clockwise or an anticlockwise closed-circuit configuration (see the actual results for 40\,nm  diameter rods shown in Fig.\,\ref{structure}c,d,e). The sequence of the out-of-plane polarization components in the four quadrants of the nanorod depends on the initial conditions. Three basic cases were found: the up-up-down-down-type domain structure with 109- and 71-degree domain boundaries (Fig.\,\ref{structure}c), the up-down-up-down-type domain structure with two 109-degree domain boundaries (Fig.\,\ref{structure}d) and the domain structure with two 71-degree domain boundaries and uniform sense of the $z$-component of the polarization (Fig.\,1e). Note that the asymmetric domain structure in Fig.\,1c contains a  71-degree boundary perpendicular to the $y$-axis and a 109-degree boundary perpendicular to the to $x$-axis, while there are only 109 degree boundaries in the $S_4(\bar{4}$) domain structure shown in Fig.\,1d.

% here  Figure 2 - switching
\begin{center}
\begin{figure}
\includegraphics[width=6.5cm]{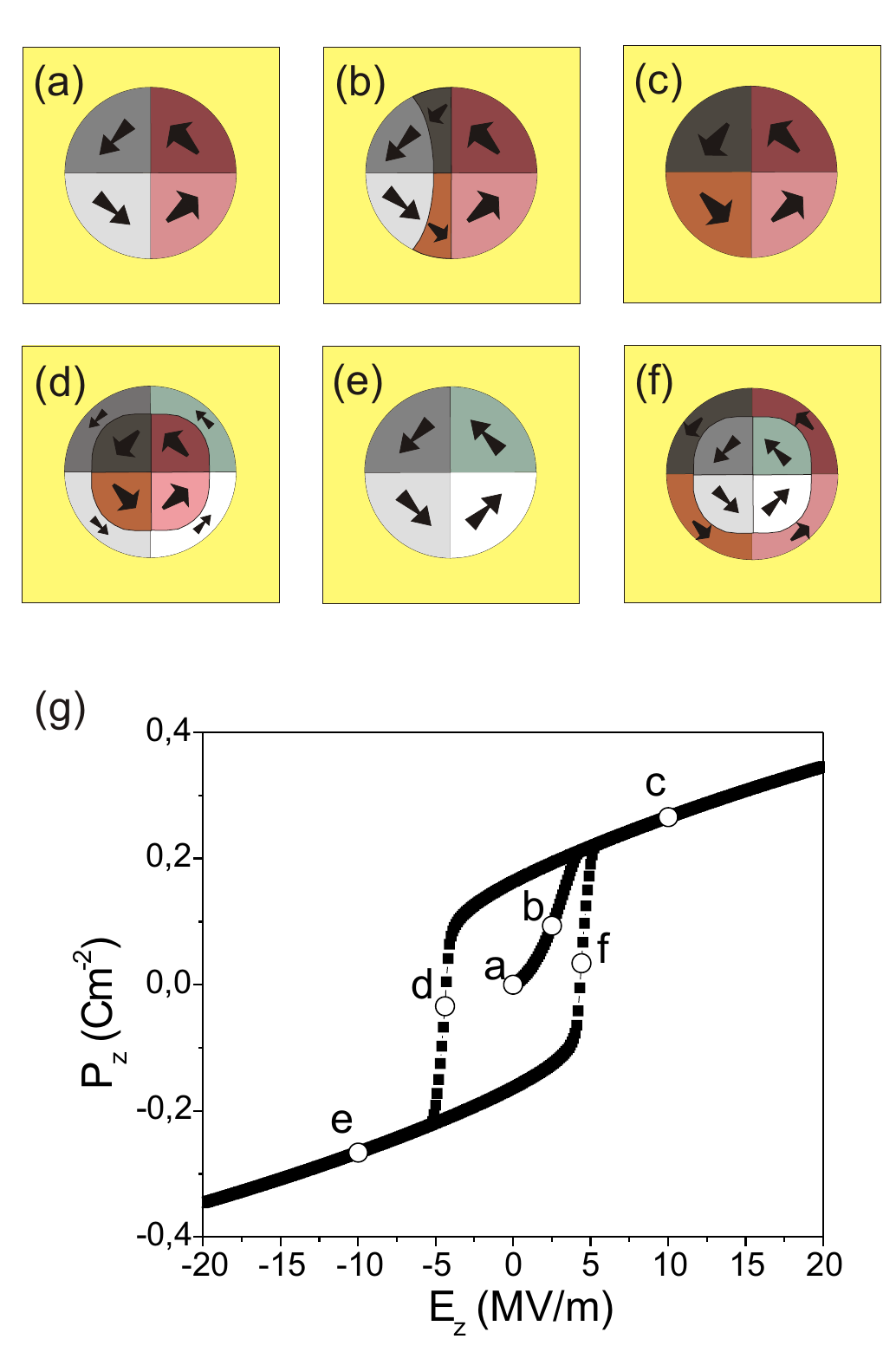}
\caption{(Color online) Evolution of the nanodomain structure within 20\,nm nanorod during the simulated switching cycle driven by electric field applied along the $z$-direction at room temperature.  Panels (a)-(f) show domain structure at selected subsequent stages of the process, panel (g) shows $P_z$ vs $E_z$ hysteresis loop, letter symbols indicate the correspondence with panels (a)-(f).
} \label{switching}
\end{figure}
\end{center}

All these domain structures are stable  with respect to a small external electric fields. Interestingly, the $P_z$ components can be reversed by  the out-of-plane electric field ${\bf E} \parallel {\bf z} $ of the order of 5\,MV/m without destruction of the closed-circuit domain quadruplet arrangement of the in-plane polarization components. Different stages of the process of polarization reversal as calculated for a 20\,nm diameter nanorod are depicted in Fig.\,2. Poling of the nanorod in a virgin state   starts by splitting of the 109-degree domain boundary into a pair of 71-degree domain boundaries, one fixed at the original position and the other one moving away (see Fig.\,2b). The moving boundary is eventually fully forced out from the nanorod when the electric field reaches about 5\,MV/m.   In the final state, the out-of-plane components of the polarization are uniformly oriented so that the total $P_z$  is maximized (see Fig.\,2c and Fig.\,2g). This configuration remains stable up to about 1\,GV/m as well as after the field removal.
In case of initial state of Fig.\,1d,  the mobile 71-degree domain boundaries peal off from both 109-domain boundaries, but otherwise the poling process is very similar to that of Fig.\,2.

The reversal of the previously poled state with the remnant polarization proceeds differently - it is realized by a collapse of a single quasi-cylindrical 71-degree domain boundary formed at the circumference surface of the nanorod (see Fig.\,2d).
This process  occurs in a fairly narrow electric field range so that the quasistatic hysteresis loop has a well defined intrinsic coercive field of about 5\,MV/m (see Fig.\,2g). The overall remnant polarization  of the nanorod (0.16\,C/m$^2$) is comparable to the spontaneous polarization of the bulk BaTiO$_3$ (0.26\,C/m$^2$ in the present model).  As expected, it decreases with temperature and vanishes near $T_{\rm C} \sim 380$\,K, which is about 10\,K below the cubic-tetragonal phase transition of stress-free bulk state (see Fig.\,3). Interestingly, the speed of the polarization reversal is limited only by the nanorod radius and domain boundary mobility and so it could be extremely fast. As an extreme example, we have seen that within the present model and its very simple kinetics, 20\,nm nanorod can be  switched by 6\,MV/m electric field within few ps.

% here  Figure 3 - teplota
\begin{center}
\begin{figure}
\includegraphics[width=6cm]{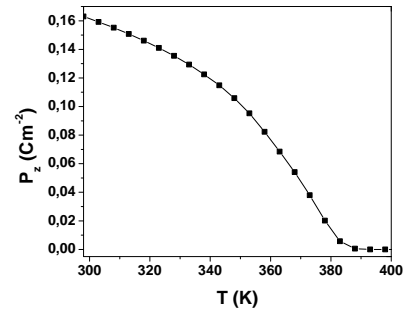}
\caption{Temperature dependence of the average remnant polarization  within the
  20\,nm BaTiO$_3$ nanorod (in the up-up-up-up state, as shown in Fig.\,1e).
} \label{teplota}
\end{figure}
\end{center}

Let us stress that in the range of stability of the rhombohedral phase, the  quadruplet states could be considered as the basic stable configuration for a broad range of diameters (about 10-80\,nm).
In our calculations the polarized domain quadruplet state is the true ground state for a 20\,nm nanorod, the asymmetric state of Fig.\,1c is the ground state for 25-70\,nm nanorod and, at 80\,nm, a multidomain  configuration is already energetically more favorable than the  quadruplet structure (see Fig.\,4).
 The overall trend shown in Fig.\,4 is probably quite generic even though the energy differences between symmetrically inequivalent quadruplets are very small and may not be experimentally relevant.
  Nevertheless, when any of these  structures is poled to the quadruplet state of Fig.\,1e, it remains stable. This is the essential property that does have a potential for practical devices. Obviously, a very thin nanorods eventually do not have domains at all; in fact, in present model the nanorods with diameter below about 10\,nm show a  tetragonal monodomain state (with the polarization oriented along the nanorod axis, see Fig.\,4).

Finally, it is worth to note that the  above discussed poled quadruplet state  has a $C_4$ macroscopic symmetry, which is a chiral one. The in-plane clockwise or anticlockwise component of the polarization could be reversed by inhomogeneous fields. Therefore, the in-plane component could be also used to store  information.
Moreover, we have noticed that the ferroelectric nanorods can also sustain a more complex domain state with the coexisting clockwise or anticlockwise layers (see Fig.\,5). In this case, the interim layer is formed by strongly bend 71-degree domain boundaries. The  core of the curling polarization is also curved and it is terminated at the circumference surface of the nanorod. Vertical motion of this interim layer could facilitate the eventual switching between the clockwise or anticlockwise state. In either case, the interim layer is preserved during the switching of the out-of-plane polarization, and it does not have any noticeable influence on the $P_z$-switching process -- the coercive field is practically identical to that of the basic $C_4$ structure.

% here  Figure 4   size-effect
\begin{center}
\begin{figure}
\includegraphics[width=8cm]{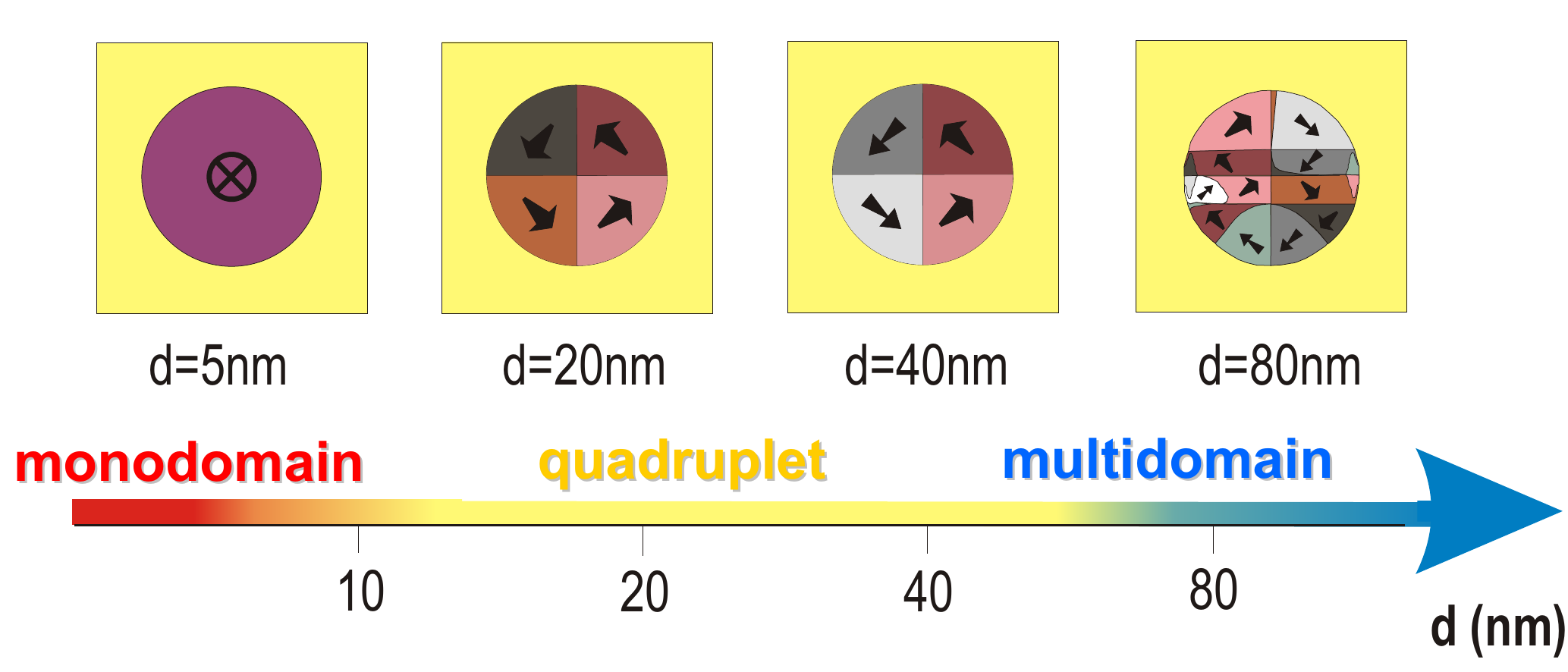}
\caption{(Color online) Evolution of the domain structure with increasing nanorod diameter (details in the main text).
} \label{size-effect}
\end{figure}
\end{center}

Interestingly, the 71-degree domain boundaries in all the above closed-circuit domain quadruplet structures  have an unusual $ \langle 100 \rangle$ orientation, which does fulfill the condition of polarization charge compatibility but which does not satisfy the conventional condition of the bulk spontaneous strain compatibility \cite{Janovec}. The same obviously also holds for the curved 71-degree domain boundaries shown in Fig.\,2 and Fig.\,5. Full discussion of this finding is beyond the scope of the present work but it is clear that it testifies a very specific softness of such boundaries and it is quite possible that this could be a typical and important property of 71-degree domain boundaries in other rhombohedral ferroelectric perovskites.

In summary, our phase-field simulations suggest that 1-3 type nanocomposites with ferroelectric nanorods enables to stabilize the interesting closed-circuit domain quadruplets. The rhombohedral closed-circuit domain quadruplets of the BaTiO$_3$/SrTiO$_3$ nanocomposite films  described here are particularly stable because the monodomain tetragonal state of the nanorod is suppressed by mechanical clamping and the rhombohedral monodomain state of the nanorod is suppressed by the electrical boundary conditions on its circumference surface. Moreover, the out-of-plane polarization of these nanorods is facilitated by nonconventional 71-degree domain boundaries that can be easily bend and nucleated from residual 109-degree domain boundaries and at the nanorod circumference surfaces.
We believe that these intriguing findings will be an inspiration for the continuation of nanoferroelectric studies.

% here  Figure 5    interim-layer
\begin{center}
\begin{figure}
\includegraphics[width=7cm]{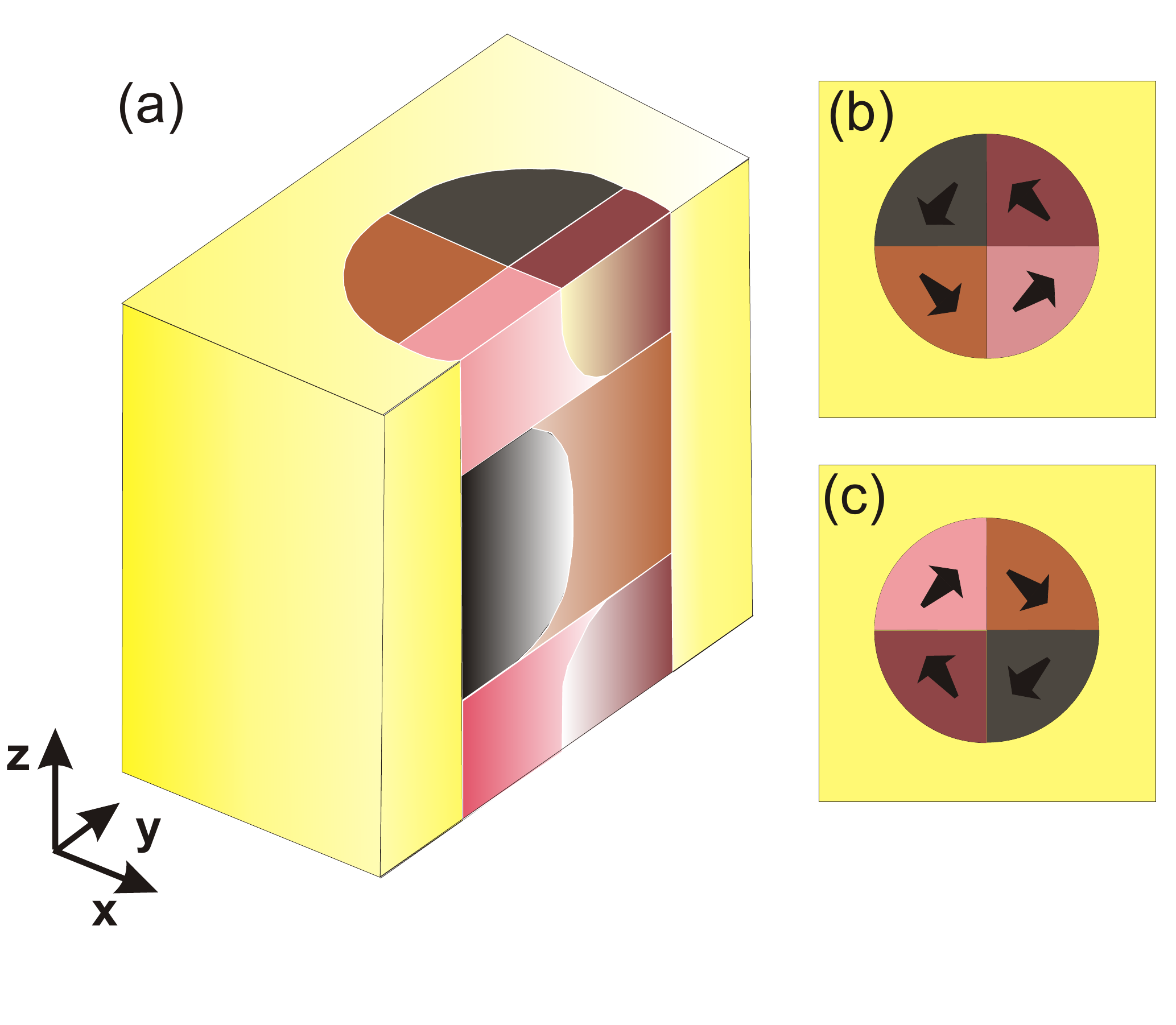}
\caption{(Color online) Domain structure with coexisting clockwise and anticlockwise curling polarization layers in a 20\,nm diameter BaTiO$_3$ nanorod: a) the overall view, b) quadruplet domain structure  as seen from the top of the displayed simulation box ($z=32$\,nm) and c) domain structure at the $z=16$\,nm cut.} \label{interim-layer}
\end{figure}
\end{center}

\begin{acknowledgments}
This work is supported by the Sciex-NMS$^{ch}$ program of the Swiss Federal Government (Project code 11.117). Additional funding was received from the European
Research Council under the EU 7th Framework Programme (FP7/2007-2013)/ERC grant
agreement no (268058) Mobile-W and by Czech Science Foundation (Project GACR P204/11/1011).
\end{acknowledgments}

%--------------------------------------------------------------------------


\begin{thebibliography}{99}

\bibitem{Scott06Rev}
J. F. Scott,
 J. Phys.: Condens. Matter {\bf 18}, R361 (2006).

\bibitem{Lee}
W. Lee, H. Han, A. Lotnyk, M.A. Schubert, S. Senz, M. Alexe, D. Hesse, S. Baik, and U. G\"{o}sele,
 Nature Nanotechnology {\bf 3}, 402 (2008).

\bibitem{Slutsker2008}
J. Slutsker, A. Artemev, and A. Roytburd,
Phys. Rev. Lett. {\bf 100}, 087602 (2008).

\bibitem{Cata13}
G. Catalan,  J. Seidel, R. Ramesh, and J. F. Scott,
 Rev. Mod. Phys. {\bf 84}, 119 (2012).

\bibitem{Mao}
Y. Mao, S. Banerjee, and S. S. Wong,
  J. Am. Chem. Soc. {\bf 125}, 15718 (2003).

\bibitem{Urban}
J. J. Urban, W. S. Yun, Q. Gu, and H. Park,
J. Am. Chem. Soc. {\bf 124}, 1186 (2002).

\bibitem{Luo2003}
Y. Luo, I. Szafraniak, V. Nagarajan, R. B. Wehrspohn, M. Steinhart, J. H. Wendorff,
N. D. Zakharov, R. Ramesh, and M. Alexe,
Integrated Ferroelectrics {\bf 59}, 1513 (2003).

\bibitem{Yun2002}
W. S. Yun, J. J. Urban, Q. Gu and H. Park,
Nano Lett. {\bf 2}, 447 (2002).

\bibitem{Saeterli2010}
R. S{\ae}terli, P. M. R{\o}rvik, C. C. You, R. Holmestad, T. Tybell, T. Grande, A. T. G. van Helvoort and M.-A. Einarsrud,
 J. Appl. Phys. {\bf 108}, 124320 (2010).

\bibitem{Vasc05A}
 E. Vasco, A. Magrez, L. Forr\'{o}, and N. Setter,
J. Phys. Chem. B {\bf 109}, 14331 (2005).

\bibitem{Wang12B}
J. Wang, A. Durussel, C. S. Sandu, M. G. Sahini, Z. He, and N. Setter,
J. Crystal Growth {\bf 347}, 1 (2012).

\bibitem{Spanier2006}
J. E. Spanier, A. M. Kolpak, J. J. Urban, I. Grinberg, L. Ouyang, W. S. Yun, A. M. Rappe, and H. Park,
 Nano Lett. {\bf 6}, 735 (2006).

\bibitem{Suya04D}
G. Suyal, E. Colla, R. Gysel, M. Cantoni, and N. Setter,
Nano Lett. {\bf 4}, 1339 (2004).

\bibitem{Wang08F}
 J. Wang, C. Stampfer, C. Roman, W. H. Ma, N. Setter, and C. Hierold,
 Appl. Phys. Lett. {\bf 93}, 223101 (2008).

\bibitem{Yama10H}
 T. Yamada, J. Wang, O. Sakata, H. Tanaka, Y. Ehara, S. Yasui, N. Setter, and H. Funakubo,
 Jpn. J. Appl. Phys. {\bf 49}, 09MC09 (2010).

\bibitem{ZWang}
Z. Wang, A. P. Suryavanshi, and M. F. Yu,
 Appl. Phys. Lett. {\bf 89}, 082903 (2006).

\bibitem{Schlom}
D. G. Schlom, L.-Q. Chen, C.-B. Eom, K. M. Rabe, S. K. Streiffer, and J.-M. Triscone, Annu. Rev. Mater. Res. {\bf 37}, 589 (2007).

\bibitem{Waru}
M. P. Warusawithana, E. V. Colla, J. N. Eckstein, and M. B. Weissman, Phys. Rev. Lett. {\bf 90}, 036802 (2003).

\bibitem{Lee05}
 H. N. Lee, H. M. Christen, M. F. Chisholm, C. M. Rouleau, and D. H. Lowndes, Nature {\bf 433}, 395 (2005).

\bibitem{Slutsker2006}
J. Slutsker, I. Levin, J. Li, A. Artemev, and A. L. Roytburd,
Phys. Rev. B {\bf 73}, 184127 (2006).

\bibitem{Yamada2009}
T. Yamada, C. S. Sandu, M. Gureev, V. O. Sherman, A. Noeth, P. Muralt, A. K. Tagantsev, and N. Setter,
 Adv. Mater. {\bf 21}, 1363 (2009).

\bibitem{Li}
J. Li, B. Nagaraj, H. Liang, W. Cao, C. H. Lee, and R. Ramesh,
Appl. Phys. Lett. {\bf 84}, 1174 (2004).

\bibitem{Gruverman2008A}
A. Gruverman, D. Wu and J. F. Scott,
 Phys. Rev. Lett. {\bf 100}, 097601 (2008).

\bibitem{Gruverman2008}
A. Gruverman, D. Wu, H.-J. Fan, I. Vrejoiu, M. Alexe, R. J. Harrison, and J. F. Scott,
J. Phys.:Condens. Matter {\bf 20}, 342201 (2008).

\bibitem{Gruverman2009}
A. Gruverman,
 J. Mater. Sci. {\bf 44}, 5182 (2009).

\bibitem{Jung}
D. J. Jung, K. Kim, and J. F. Scott,
J. Phys: Condens. Matter {\bf 17}, 4843 (2005).

\bibitem{Fridkin}
V. M. Fridkin, R. V. Gaynutdinov and S. Ducharme,
 Uspekhi Fizicheskikh Nauk {\bf 180 (2)}, 209 (2010).

\bibitem{Kim}
Y. Kim, H. Han, W. Lee, S. Baik, D. Hesse, and M. Alexe,
Nano Lett. {\bf 10}, 1266 (2010).



\bibitem{Wang2011}
J. J. Wang, E. A. Eliseev, X. Q. Ma, P. P. Wu, A. N. Morozovska, and L.-Q. Chen,
  Acta Materialia {\bf 59}, 7189 (2011).

\bibitem{Ondr13}
P. Ondrejkovic, P. Marton, M. Guennou, N. Setter, and J. Hlinka,
Phys. Rev. B {\bf 88}, 024114 (2013).

\bibitem{Schilling2006}
A. Schilling, T. B. Adams, R. M. Bowman, J. M. Gregg, G. Catalan, and J. F. Scott,
 Phys. Rev. B. {\bf 74}, 024115 (2006).

\bibitem{Schilling2009}
A. Schilling, D. Byrne, G. Catalan, K. G. Webber, Y. A. Genenko, G. S. Wu, J. F. Scott, and J. M. Gregg,
Nano Lett. {\bf 9}, 3359 (2009).

\bibitem{Fu}
H. Fu and L. Bellaiche,
Phys. Rev. Lett. {\bf 91} 257601 (2003).

\bibitem{Naumov}
I. I. Naumov, L. Bellaiche, and H. Fu,
Nature {\bf 432}, 737 (2004).

\bibitem{Stachiotti}
M. G. Stachiotti and M. Sepliarsky,
 Ferroelectrics {\bf 427}, 41 (2012).

\bibitem{Baudry}
L. Baudry, I. A. Luk'yanchuk and A. Sen\'{e},
 Ferroelectrics {\bf 427}, 34 (2012).

\bibitem{Baudry2012}
L. Baudry, I. A. Luk'yanchuk, and A. Sen\'{e},
Integrated Ferroelectrics {\bf 133}, 96 (2012).

\bibitem{Chen2012}
W. J. Chen, Y. Zheng, and B. Wang,
Appl. Phys. Lett. {\bf 100}, 062901 (2012).

\bibitem{Laho08}
L. Lahoche, I. A. Luk'yanchuk, and G. Pascoli, Integrated Ferroelectrics {\bf 199}, 60 (2008).

\bibitem{Schilling}
A. Schilling, S. Prosandeev, R. G. P. McQuaid, L. Bellaiche, J. F. Scott, and J. M. Gregg,
Phys. Rev. B {\bf 84}, 064110 (2011).

\bibitem{Gregg}
J. M. Gregg,
Ferroelectrics {\bf 433}, 74 (2012).

\bibitem{Boro13}
F. Borodavka, I. Gregora, A. Bartasyte, S. Margueron, V. Plausinaitiene, A. Abrutis, and J. Hlinka, J. Appl. Phys. {\bf 113}, 187216 (2013).

\bibitem{Ahlu13}
R. Ahluwalia, N. Ng, A. Schilling, R. G. P. McQuaid, D. M. Evans, J. M. Gregg, D. J. Srolovitz, and J. F. Scott,
Phys. Rev. Lett. {\bf 111}, 165702 (2013).

\bibitem{McQuaid}
R. G. P. McQuaid, L. J. McGilly, P. Sharma, A. Gruverman, and J. M. Gregg,
 Nature Communications {\bf2}, 404 (2011).

\bibitem{Chang}
L.-W. Chang, V. Nagarajan, J. F. Scott and J. M. Gregg,
Nano Lett. {\bf 13}, 2553 (2013).

\bibitem{Ivry}
Y. Ivry, D. P. Chu, J. F. Scott, and C. Durkan,
 Phys. Rev. Lett. {\bf 104}, 207602 (2010).

\bibitem{McGilly2010}
L. J. McGilly, A. Schilling, and J. M. Gregg,
 Nano Lett. {\bf 10}, 4200 (2010).

\bibitem{my}
 J. Hlinka and P. Marton, Phys. Rev. B {\bf 74}, 104104 (2006).

\bibitem{Hlin09}
 J. Hlinka, P. Ondrejkovic, and P. Marton, Nanotechnology {\bf 20}, 105709 (2009).

\bibitem{Mart10}
P. Marton, I. Rychetsky, and J. Hlinka, Phys. Rev. B {\bf 81}, 144125 (2010).

\bibitem{HuChen3D}
 H.-L. Hu and L.-Q. Chen, J. Am. Ceram. Soc. {\bf 81}, 492 (1998).

\bibitem{Khacha}
 S. Semenovskaya and A. G. Khachaturyan, J. Appl. Phys. {\bf 83}, 5125 (1998).

\bibitem{Nambu}
 S. Nambu and D. A. Sagala, Phys. Rev. B {\bf 50}, 5838 (1994).

\bibitem{Artemev}
 A. Artemev, J. Slutsker, and A. L. Roytburd, IEEE Trans. Ultras. Ferr. Freq. Contr. {\bf 55}, 963 (2008).

\bibitem{simul}
 P. Marton and J. Hlinka, Phase Transitions {\bf 79}, 467 (2006).

\bibitem{Klot03}
J. Hlinka and E. Klotins,  J. Phys.: Condens. Matter {\bf 15}, 5755 (2003).

\bibitem{Sheng}
G. Sheng, Y. L. Li, J. X. Zhang, S. Choudhury, Q. X. Jia, V. Gopalan, D. G. Schlom, Z. K. Liu, and L. Q. Chen,
 Appl. Phys. Lett. {\bf 96},  232902, (2010).

\bibitem{Yama72} T. Yamada, J. Appl. Phys. {\bf 43} 328 (1972).

\bibitem{timescale}
 J. Hlinka, Ferroelectrics {\bf 349}, 49 (2007).

\bibitem{Janovec}
J. Fousek and V. Janovec,
J. Appl. Phys. {\bf 40}, 135 (1969).


\end{thebibliography}
\end{document}